\documentclass[twocolumn,showpacs,preprintnumbers,amsmath,amssymb]{revtex4}

\usepackage{graphicx}
\usepackage{dcolumn}
\usepackage{bm}

\begin{document}

\title{"Equilibrium" states of nonequilibrium system.}

\author{Bohdan Lev}

\affiliation{Department of the Synergetic, Bogolyubov Institute
of the Theoretical Physics, NAS Ukraine, Metrologichna
14-b,03680, Kyiv, Ukraine.}

\date{\today}

\begin{abstract}

A possible approach to description of the nonequilibrium system
has been proposed. Based on the Fokker-Plank equation in term of
energy for nonequilibrium distribution function of macroscopical
system was obtained the stationary solution which can be
interpreted as the equilibrium distribution function for new
energetic state. The proposed approach takes into account the
possible motion between different states of system, induced by
dissipation of energy and influence of environment which
dependence on energy of the system. A non-linear model, which
describe possible stationary state of system with different
processes in it, has been described.

\end{abstract}
\maketitle

In the most cases, any macroscopical system is nonequilibrium.
Properties of such system determined by the internal condition and
connection with environment \cite{Bal},\cite{Huang}. Equilibrium
distribution function of a system can be determined for very
specific conditions \cite{Huang},\cite{Zub}. The transition of the
system from equilibrium to nonequilibrium is the result of the
external influence on system which change the initial state of
the system. There are no well-defined methods of definition of
nonequilibrium distribution function, which can take into account
the possible states of the system. Standard method allow to
expresses nonequilibrium states based on the information about
equilibrium state and small deviation from this state. The
well-known fluctuation-dissipation theorem defines the possible
small fluctuation around equilibrium state. The nonequilibrium in
this approach stand out in form of small modifications of the
equilibrium distribution function and relaxed to equilibrium
state. The main questions in nonequilibrium system description
can be formulated next: \textit{How the nonequilibrium state can
be describe in general case and what is the condition of it's
transformation to the equilibrium state? Is it possible to
determine the new state of the system taking to account the
dissipation and influence of the environment? What is this state
and how this state can be characterized taking into account the
possible dissipation in the system and influence of the
environment?}

Every system in equilibrium state can be described in the phase
space in the term distribution function. The distribution
functions which determine all thermodynamic properties of
macroscopic system depends on the energy of this system. In
practice, detailed knowledge of only few macroscopic parameters
such as energy is necessary and that is enough for statistical
description of the system. It is not possible to watch up on
detail microscopical dynamic on the phase space, but it is
possible to determine the energy as control parameter of the
system. In the absence of any other knowledge about the system,
there is no reason to favor any state of system determined trough
energy.  A change of energy of the system determines a change of
the state of the macroscopical one. The nonequilibrium
distribution function in general case depended on energy of
system. In case of nonequilibrium the energy is changing and
system can transit from one state to another. It depend on
external influence and initial peculiarity. The external
influence, first of all, manifests in change of energy of the
system. This changing of energy of system can be taken into
account as random function of time. That is result of influence
of all other macroscopical system which present the environment
and from which determine behaviour of single system. Also, the
energy of single system changed if it dissipate or absorb the
energy. The nonequilibrium distribution function as in
equilibrium case can define as $\rho(E,t)$, which include the
dependence on energy of the system $E$ and time. The evolution of
system can be expressed in terms of distribution function only,
which takes to account the dissipation energy and a random
migration along the energy. The random migration of the system is
the result of interaction between this system and the
environment, the influence of which should prove the random
change of the energy of the system. The system which can not rich
the equilibrium after fast change of the environment must
relaxing to the new state. This process takes to account possible
degradation of the system in contact with environment.
Furthermore, the nonequilibrium fluctuation in the system can be
described and probability that the new state of system arise can
be determined. One can introduce the variable which describe the
change of energy of system as $\varepsilon=E-E_{0}$, where $E_{0}$
correspond to the initial state which can be equilibrium. The
general equation which defines the change of energy and takes to
account the dissipation energy and random walk along energy can be
expressed in the form:
\begin{equation}
\frac{d \varepsilon}{d t}=f(\varepsilon)+L(t)
\end{equation}
where $f(\varepsilon)$ defines the function of energy change at
different dissipation processes allowed within current system,
and $L(t)$ determines possible fluctuation, which can not
described by previous part and includes the random disturbance
from the environment. The dependence of dissipation function on
the energy describes all possible processes under the external
influence. The different interesting cases of dependence of this
function on energy will be shown below. It is important to note,
that in any case this function must depend on initial energy,
because the system does not know the direction of it's
transformation. The correlation between two values of fluctuation
$L(t)$ at two different moments is not zero only for time
interval, which is equal to the time of action. It is possible to
explain by a relation $\left\langle
L(t_{1})L(t_{2})\right\rangle=\phi(t_{1}-t_{2})$. The function
$\phi \delta(t_{1}-t_{2})$ must have the drastic peak at environs
of zero \cite{Kam},\cite{Hor} and satisfy the condition $\int
\phi(\tau)d\tau=\sigma^{2}$ for the white noise. The present
equation for energy lead to Fokker-Plank equation for
nonequilibrium distribution function which can be write in the
form:
\begin{equation}
\frac{\partial \rho(\varepsilon,t)}{\partial
t}=-\frac{\partial}{\partial
\varepsilon}\left(f(\varepsilon)\rho(\varepsilon,t)\right)+\frac{\sigma^{2}}{2}\frac{\partial^{2}
\rho(\varepsilon,t)}{\partial \varepsilon^{2}}
\end{equation}
This equation can be reexpressed in form of local the distribution
function conservation law:
\begin{equation}
\frac{\partial \rho(\varepsilon,t)}{\partial t}=\frac{\partial
J(\varepsilon)}{\partial \varepsilon}
\end{equation}
where flux of distribution function is:
\begin{equation}
J(\varepsilon)=-f(\varepsilon)\rho(\varepsilon,t)+\frac{\sigma^{2}}{2}\frac{\partial
\rho(\varepsilon,t)}{\partial \varepsilon}
\end{equation}
The stationary solution of this equation for absence the flow of
probability $J(\varepsilon)=0$ can obtain in the simple form:
\begin{equation}
\rho(\varepsilon)=A
exp\left(\int^{\varepsilon}_{0}\frac{f(\varepsilon')}{\sigma^{2}}d\varepsilon'\right)
\end{equation}
Must be noted that, $\varepsilon=0$ is not only inner limit but
stationary point at the absence of the dissipation energy for
conservative system $f(\varepsilon)=0$ and random diffusion. This
point, simultaneously, is the attractive point and thus whole
stationary probability must be concentrate at zero, according the
normalization condition for distribution function \cite{Hor}.
Only in this case distribution function can correspond the
microcanonical distribution function.

Furthermore, for present only random diffusion on energy the
equation for the nonequilibrium distribution function take the
form of diffusion equation and solution is:
\begin{equation}
\rho(\varepsilon)=A \frac{1}{\sqrt{4\pi \sigma^{2}
t}}exp\left(-\frac{ \varepsilon^{2}}{4\sigma^{2}t}\right)
\end{equation}
that describe migration of the system along energy. The measure of
blueness increases in time according to the law $\left\langle
\varepsilon^{2}\right\rangle=2\sigma^{2}t$. This solution
represent evolution of the system which in initial state
described by the equilibrium distribution function
$\rho(\varepsilon)=\delta(E-E_{0})$. All states of system at the
initial moment are in the point on the energy conservation
surface. The fluctuation of external media manifests in absence of
the microcanonical distribution and presence of the uniform
distribution on all energies. In the case, when the dissipation
of energy can be represented as function $f(\varepsilon)=-\gamma
\varepsilon$ which accounts for possible friction of the system
inside external medium, the stationary solution can be rewritten
in the another form:
\begin{equation}
\rho(\varepsilon)=A exp\left(-\frac{\gamma
\varepsilon^{2}}{\sigma^{2}}\right)
\end{equation}
For determination of physical sense of coefficient of
distribution must return to dynamic equation for energy. The
solution of Langevine equation can be present in the form:
\begin{equation}
\left\langle \varepsilon^{2}\right\rangle
=\varepsilon^{2}_{0}exp(-2\gamma
t)+\frac{\sigma^{2}}{2\gamma}\left(1-exp(-2\gamma t)\right)
\end{equation}
with necessary condition $\lim_{t\rightarrow \infty}\left\langle
\varepsilon^{2}\right\rangle=\frac{\sigma^{2}}{2\gamma}$. The
nonequilibrium distribution function can be rewritten as:
\begin{equation}
\rho(\varepsilon)=A exp\left(-\frac{\varepsilon^{2}}{\left\langle
\varepsilon^{2}\right\rangle}\right)
\end{equation}
which does look like the equilibrium Gaussian distribution
function for the energy near average value. Apparently from
previous statement, in simple case of random diffusion the
equilibrium distribution function can not take the well-know form.
However, there is a need in further consideration of most general
case, when the system is in a contact with a nonlinear external
medium. The general dissipation equation can be written as
\cite{Kam},\cite{Hor}:
\begin{equation}
\frac{d \varepsilon}{d t}=f(\varepsilon)+g(\varepsilon)L(t)
\end{equation}
where the second part describe the nonlinearity of diffusion
process. In this general case the equilibrium ratio of fluctuation
of the energy can presented as \cite{Lan}
\begin{equation}
\left\langle
\varepsilon^{2}\right\rangle=\frac{\alpha(0)}{\beta}=\frac{\sigma^{2}}{2}g^{2}(0)
\end{equation}
where $\alpha(0)$ is sensitivity of system,$\alpha(0)=c_{v}$ is a
heat and $\beta$ is inverse temperature of the system at
equilibrium. For the nonlinear and nonequilibrium state, the
general representation of sensitivity of system for different
energy can be introduced in the form:
\begin{equation}
\alpha(\varepsilon)=\frac{\beta \sigma^{2}}{2}g^{2}(\varepsilon)
\end{equation}
where possible dependence of reaction of the system on external
influence is included.

For the true nonlinear Langevine equation must exist equivalent
equation for probability distribution function which can wrote,
according to physical process. In this time has been propose two
different approaches. If one consider the dependence of the
coefficient $g(\varepsilon)$ only on energy at start point, the
equation for nonequilibrium distribution function can be obtained
in the Ito form. Also, if this coefficient did depend on energy
before and after jump, the diffusive equation can be written in
the Stratonovich form:
\begin{equation}
\frac{\partial \rho(\varepsilon,t)}{\partial
t}=-\frac{\partial}{\partial
\varepsilon}\left(f(\varepsilon)\rho(\varepsilon,t)\right)+\frac{\sigma^{2}}{2}\frac{\partial}{\partial
\varepsilon }g(\varepsilon)\frac{\partial}{\partial
\varepsilon}g(\varepsilon)\rho(\varepsilon,t)
\end{equation}
Below only Statonovich presentation is used, because both
presentation are connected \cite{Hor}, \cite{Kam}. Both equations
in different form does not have much sense. The physical
interpretation of the processes must be studied. In described
case, representation of different states of every system
determined and can be formed by previous state and possible
future states. The present equation for nonequilibrium
distribution function in this case can be reexpressed in more
usual form for local conservation law for probability:
\begin{equation}
\frac{\partial \rho(\varepsilon,t)}{\partial
t}=\frac{\partial}{\partial \varepsilon}J(\rho(\varepsilon,t))
\end{equation}
The flow of probability can be written as:
\begin{equation}
J=-\left(f(\varepsilon)-\frac{\sigma^{2}}{2}g(\varepsilon)\frac{\partial}{\partial
\varepsilon }g(\varepsilon)
\right)\rho(\varepsilon,t)+\frac{\sigma^{2}}{2}g^{2}(\varepsilon)\frac{\partial}{\partial
\varepsilon}\rho(\varepsilon,t)
\end{equation}
The stationary solution for $J(\rho(\varepsilon,t))=0$, without
flow of probability trough every point can presented by:
\begin{equation}
\rho_{s}(\varepsilon)=A exp\left\{\int^{\varepsilon}_{0}\frac{2
f(\varepsilon')d\varepsilon'}{\sigma^{2}g^{2}(\varepsilon')}-\ln
g(\varepsilon)\right\}
\end{equation}
The stationary solution can interpretation as equilibrium
distribution function in present state. The physical conditions
and peculiarity of the interaction of the system with the
external medium must be taken under consideration in order to
take into account the degradation processes in the system. The
equilibrium distribution function as stationary solution can be
presented as:
\begin{equation}
\rho_{s}(\varepsilon)=A exp\left\{-U(\varepsilon)\right\}
\end{equation}
where
\begin{equation}
U(\varepsilon)=\ln g(\varepsilon)-\int^{\varepsilon}_{0}\frac{2
f(\varepsilon')d\varepsilon'}{\sigma^{2}g^{2}(\varepsilon')}
\end{equation}
The extremal value of this distribution function can be obtained from the
equation
\begin{equation}
U'(\widetilde{\varepsilon})=\frac{1}{D(\varepsilon)}\left(D'(\varepsilon)-f(\varepsilon)\right)
\end{equation}
where $'$ stands for the energy derivative. This equation is
equal to
\begin{equation}
D'(\widetilde{\varepsilon})=f(\widetilde{\varepsilon})
\end{equation}
which determines the relation between dissipation in the system
and the diffusion along energy for stationary case and fully
determine new equilibrium state of system. The stationary
nonequilibrium distribution function can be given as
\begin{equation}
\rho_{s}(\varepsilon)=exp\left\{-U(\widetilde{\varepsilon})\right\}
exp\left(-U''(\widetilde{\varepsilon})\varepsilon^{2}\right)
\end{equation}
where
\begin{equation}
-U''(\widetilde{\varepsilon})=\frac{1}{D(\widetilde{\varepsilon})}\left(D''(\widetilde{\varepsilon})-f'(\widetilde{\varepsilon})\right)
\end{equation}
This equilibrium distribution function has a Gaussian form.

In the case of energy conservation, $f(\varepsilon)=0$, the
stationary solution became:
\begin{equation}
\rho_{s}(\varepsilon)=A exp\left\{-\ln g(\varepsilon)\right\}
\end{equation}
that corresponds to the canonical equilibrium distribution
function only if $g(\varepsilon)=e^{-\beta \varepsilon}$, where
$\beta$ is the inverse temperature. It is possible only if the
diffusion coefficient is:
\begin{equation}
D(\varepsilon)=\frac{\sigma^{2}}{2}g^{2}(\varepsilon)=\frac{\sigma^{2}}{2}e^{2\beta
\varepsilon}
\end{equation}
Taking to account the presentation of the diffusion coefficient
in this case can obtain:
\begin{equation}
-U''(\widetilde{\varepsilon})=2\beta\left(2\beta-\frac{f'(\widetilde{\varepsilon})}{f(\widetilde{\varepsilon})}\right)
\end{equation}
where $\widetilde{\varepsilon}$ is solution of the equation
\begin{equation}
exp (2\beta
\widetilde{\varepsilon})=\frac{2}{\sigma^{2}}f(\widetilde{\varepsilon})
\end{equation}
The equilibrium distribution function can be written in the form
\begin{equation}
\rho_{s}(\varepsilon)=A
exp\left\{-2\beta\left(2\beta-\frac{f'(\widetilde{\varepsilon})}{f(\widetilde{\varepsilon})}\right)\varepsilon^{2}\right\}
\end{equation}
If the dissipation of the system $f(\varepsilon)$ is represented
by the nonlinear function of the state, many interesting
situations, including the noise-induced transition in new
equilibrium states, which will be more stabile than previous
states, can be obtained. The diffusion coefficient is an universal
characteristic of the environment. The dependence of the
diffusion coefficient on energy can be determined from linear
Langevine equation by using the theory of Markovian processes,
and take into account the fluctuation of different coefficients
in function $f(\varepsilon)$. If this dissipation function
present in form $f(\varepsilon)=\alpha_{t}e^{\beta \varepsilon}$
the dissipation Langevine equation (1) can be rewritten in the
other form
\begin{equation}
\frac{d e^{-\beta \varepsilon(t)}}{d t}=-\beta \alpha_{t}
\end{equation}
where $\alpha_{t}=\alpha +\xi_{t}$ has a constant part and a part
$\xi_{t}$ which describe the influence of the white noise of the
environment \cite{Hor}. The Fokker-Plank equation for the
nonequilibrium distribution function can be present in new
variable $z=e^{-\beta \varepsilon}$, in another simple form
\begin{equation}
\frac{\partial \rho(z,t)}{\partial t}=\frac{\partial}{\partial
z}(\alpha \beta
\rho(z,t))+\frac{\sigma^{2}\beta^{2}}{2}\frac{\partial^{2}}{\partial
z^{2}}\rho(z,t)
\end{equation}
which has the stationary solution
\begin{equation}
\rho_{s}(z)=exp(\frac{2\alpha}{\sigma^{2}\beta} z)
\end{equation}
In case $\beta \varepsilon > 1$, one can obtain
\begin{equation}
\rho_{s}(\varepsilon)= exp\left\{-\beta \varepsilon\right\}
\end{equation}
for ${\frac{2 \alpha}{\sigma^{2}\beta}}=1$. This solution is equal
to the solution which was obtained previously by standard
approach.

The energetic presentation of description the nonequilibrium
process is valid only in case if this variable is canonical and
carry out averaging on phase as conjugate value. But this
presentation can be more illustrative if take an interest in
ordinary way to description equilibrium states. This conception
cab be illustrate on the example usual Brownian particle.The
motion of Brownian particle usually describe in impulse space
while coordinate determined as position of particle in result
many random jump of particle. The energy of particle can be
characterization only in term kinetic energy
$\varepsilon=\frac{p^{2}}{2M}$, in previous presentation
$E_{0}=0$. The usually Langevine equation can write in the form
\begin{equation}
\frac{d p}{d t}=-\gamma p +F(t)
\end{equation}
where $F(t)$ is random force. Using this relation and
presentation of energy of Brownian particle can write the
Langevine equation  in the form
\begin{equation}
\frac{d \varepsilon}{d t}=\frac{p}{M}\frac{d p}{d t}\equiv
-\gamma \varepsilon-\sqrt{\varepsilon}L(t)
\end{equation}
that completely agree with equation (10) with
$f(\varepsilon)=-2\gamma \varepsilon$ and $g(\varepsilon)=\sqrt{
\varepsilon}$. The proposed in this article approach can be write
the stationary solution in the form (16) that for the case
Brownian particle gives
\begin{equation}
\rho_{s}(\varepsilon)=A\exp\left\{-\frac{4\gamma}{\sigma^{2}}\varepsilon-\ln
\sqrt{\varepsilon} \right\}\equiv A\frac{1}{\sqrt{
\varepsilon}}\exp(-2 \beta \varepsilon)
\end{equation}
where using well-known relation between characteristic white noise
and temperature $\frac{2\gamma}{\sigma^{2}}=\beta$. If take into
account normalization condition that $\int \rho_{s}(\varepsilon)
d\varepsilon \equiv \int \rho_{s}(p) d p $ can obtain the
equilibrium distribution function in impulse space in the form
\begin{equation}
\rho_{s}(p)= A \exp\left(-2\beta \frac{p^{2}}{2M}\right)=A
\exp\left(-\beta M v^{2}\right)
\end{equation}
where $v$ is velocity of particle. Obtained stationary solution
fully represented well-known equilibrium distribution function for
Brownian particle.

In conclusion, one can consider the very simple picture about the
motion of Browning particle in the heterogeneous media. For this
matter the characteristic of the media can be take into account
by different value of the friction coefficient, which depend on
one spatial point as random quantity. As example, the motion of a
large particle in a suspension, a dusty particle in
nonhomogeneous media and some other interesting cases, when it is
necessary to determine the equilibrium distribution function of
the Browning particles for description of kinetic properties of
the matter. In this case, the general results \cite{Hor} can be
used, where present approach used to describe noise induced phase
transition. Therefore, the Langevine equation in velocity variable
can be written in the following form $\frac{d v}{d
t}=-\gamma_{t}v$, where $\gamma_{t}=\gamma +\xi_{t}$ consist of
constant part $\gamma $, which determine the average friction
coefficient and chaotic part $\xi_{t}$, which describe the
influence of random change in friction of matter. In case of
white noise of possible fluctuation of density or other parameter
which characterize matter the Fokker-Plank equation for density
of probability or the nonequilibrium distribution function in the
Stratonovich interpretation in the standard form can be written
\cite{Hor}:
\begin{equation}
\frac{\partial \rho(v,t)}{\partial t}=\frac{\partial}{\partial
v}\left(\gamma v
\rho(v,t)\right)+\frac{\sigma^{2}}{2}\frac{\partial^{2}}{\partial
v^{2} }v^{2}\rho(v,t)
\end{equation}
The stationary solution of this equation as equilibrium
distribution function is \cite{Hor}:
\begin{equation}
\rho_{s}(v,t)=N v^{-(\frac{2\gamma}{\sigma^{2}}+1)}
\end{equation}
which can be checked by direct substitution of the solution in
the previous equation. This stationary solution is different from
solution in standard case, when the diffusion coefficient and the
friction coefficient do not depended on spatial point and do not
fluctuate in the matter. Not difficult to present this result in
energetic presentation if take into account the dependence
diffusion coefficient from energy as $\gamma= \gamma(p)\sim
\varepsilon $. In this case can obtain the stationary
distribution function from energy as
\begin{equation}
\rho_{s}(\varepsilon)=N
\varepsilon^{-(\frac{\gamma}{\sigma^{2}}+1)}
\end{equation}
This result can be examine experimentally.

The possible approach to description of the nonequilibrium states
in energetic presentation has been proposed. Based on the
Fokker-Plank equation for nonequilibrium distribution function of
macroscopical system was obtained the stationary solution which
can be interpreted as the equilibrium distribution function for
new energetic state. The proposed approach takes into account the
possible motion between different states of system, induced by
dissipation of energy and influence of environment, which
dependence from energetic of system. A non-linear model, which
describe possible stationary state of system with different
spacial process in it, has been described. Is determined the
energy of new state of  system and new equilibrium distribution
function of state. All theoretical results can be examine
experimentally. Author acknowledge useful discussion with Dr
A.A.Semenov which made indication the example for Brownian
particle and S.B. Chernyshuk for the support.

\end{document}